# CHANNEL ACCESS CLIENT TOOLBOX FOR MATLAB

Andrei Terebilo, SLAC, Stanford, CA 94025, USA

Abstract

This paper reports on MATLAB Channel Access (MCA) Toolbox – MATLAB [1] interface to EPICS Channel Access (CA) client library. We are developing the toolbox for SPEAR3 accelerator controls, but it is of general use for accelerator and experimental physics applications programming. It is packaged as a MATLAB toolbox to allow easy development of complex CA client applications entirely in MATLAB. The benefits include: the ability to calculate and display parameters that use EPICS process variables as inputs, availability of MATLAB graphics tools for user interface design, and integration with the MATLAB-based accelerator modeling software - Accelerator Toolbox [2-4]. Another purpose of this paper is to propose a feasible path to a synergy between accelerator control systems and accelerator simulation codes, the idea known as on-line accelerator model.

## 1 BACKGROUND

The SPEAR3 light source [5] is going to be commissioned in 2004. The control system is undergoing major upgrades [6]. Most machine parameters will be accessible as EPICS Process Variables (PV's). This improvement allows an efficient division of software development effort between accelerator controls group and accelerator physics group. Examples of applications to be developed by the accelerator physics group are:

- Interactive orbit control - graphical drag-and-drop control of closed orbit, local beam bumps, and slow orbit feedback [7].
- Beam based alignment – experimental and model based determination of orbit offsets in quadrupoles with subsequent correction.
- Automated testing of orbit interlock – systematic testing of the closed orbit configurations designed to trip the interlock.
- Linear optics correction – experimental determination of linear optics parameters from measured corrector-to-BPM response matrix and subsequent model parameter fitting.

All of these applications use the real machine data in computations that may involve complex numeric algorithms or accelerator modeling. To implement this functionality in standalone control applications using a low-level programming language would be a laborious task.

It is now widely recognized that access to online accelerator models is a highly desirable feature in controls applications. In several accelerator facilities, functioning systems have been developed using a variety of software technologies and accelerator modeling codes [8,9].

However, the problem for the rest of us remains unsolved. The main difficulty is the absence of a simple and flexible protocol for communication between a control system and accelerator modeling code. The solution we have found for SPEAR3 is to make the real machine data, as well as accelerator physics simulations, available in a powerful computational environment, external to the control system. Our MATLAB Channel Access Toolbox is an important part of this design.

## 2 MCA TOOLBOX

### 2.1 MATLAB Toolbox philosophy

The MATLAB Channel Access Toolbox (MCA) implements most of the functionality of CA client library in a small number of MATLAB functions. These functions can be called from the MATLAB command line or used in a program written in MATLAB language. Values to be read from or to be written to PV's are variables in the MATLAB workspace. They are immediately available for use in calculations or plotting in MATLAB.

All functions in the toolbox have the standard 'help' header that can be displayed in the command window with MATLAB command HELP.

### 2.2 Opening a connection

MCA uses numeric handles to identify PV's instead of PV names. The function MCAOPEN[1] opens a connection to a PV and returns a numeric handle.

>> Handle = mcaopen('ExamplePVName');

---

[1] Function names appear in ALL CAPS in the main text of the paper. In MATLAB examples, they appear in lower case – the same way they are entered from MATLAB command prompt.

From this point on, all other functions in MCA toolbox refer to this PV by its handle. The connection is maintained independently from the user. If a server or network error occurs, the EPICS CA client library will periodically attempt to restore the connection until it succeeds or until the user explicitly closes the connection with MCACLOSE. At any time the connection state can be checked with MCASTATE. Another function MCAINFO returns additional information about a PV such as native data type, number of elements and the host name of the CA server.

### 2.3 Data types

The only numeric type that MATLAB recognizes for calculations is double precision floating point. The user does not need to know the native record type of a PV. Function MCAGET reads all numeric or enumerated types into the MATLAB workspace as double non-complex numbers or vectors.

PV's whose native EPICS type is 'string' become MATLAB strings. String PV's with more than one element are read as MATLAB cell arrays of strings.

### 2.4 Array PV's

The user does not need to specify the number of elements in array PV's to read or write. The function MCAGET reads the entire array into a MATLAB array. MCAPUT writes the values into a PV.

```
>> ArrayHandle = mcaopen('ArrayPVName');
>> ArrayData = mcaget(ArrayHandle);
>> mcaput(ArrayHandle,NewData);
```

### 2.5 Data validation

If MCAGET fails, it returns an empty MATLAB array. MCAPUT returns one on success, zero on failure. These operations can fail due to server or network problems. This can be checked with MCASTATE. Another possible reason is that the timeout is too short. The user can view or change the timeout settings for 'gets' and 'puts' with MCATIMEOUT

### 2.6 Asynchronous CA operations

MCA Toolbox allows asynchronous CA operations such as monitors. MCAMON installs a monitor on a PV. When a monitor is installed, MCA maintains a local copy of the value which updates only when the value changes on the server. Another function MCACACHE retrieves the most recent local value without any network communication.

```
>> mcamon(Handle);
>> Value = mcacache(Handle);
```

MCAMON can also install a monitor callback function that will be executed when the value changes. This callback function may, for example, redraw a MATLAB GUI element or start a calculation that uses the new PV value as input. Callback functions are also written in the MATLAB language. In the following example the callback is set to modify the value of a workspace variable.

```
>> mcamon(Handle ,'LiveValue=mcacache(Handle)');
```

Previously mentioned MCAPUT is actually implemented as an asynchronous call. If the timeout is set too short, MCAPUT may return zero (failure) but the value is still written to a PV.

### 2.7 Availability

MCA has been developed and tested on Windows platforms. It is available for download at
http://www-ssrl.slac.stanford.edu/mca/

## 3 ACKNOWLEDGEMENTS

The work reported on here was supported in part by DOE Contract DE-AC03-76SF00515 and Office of Basic Energy Sciences, Division of Chemical Sciences.